\documentclass[12pt,preprint]{aastex}
\usepackage{graphicx}
\usepackage{epsfig}
\usepackage{amsmath,amssymb}
\usepackage{lscape}
\usepackage{verbatim}

\shorttitle{Evolution in the Iron Abundance of the ICM}
\shortauthors{Anderson et al.}

\begin{document}
\title{Redshift Evolution in the Iron Abundance of the Intracluster Medium}
\author{Michael E. Anderson\altaffilmark{1}, Joel N. Bregman\altaffilmark{1}, Suzanne C. Butler\altaffilmark{1,2}, C. R. Mullis\altaffilmark{1}\\ }
\altaffiltext{1}{Department of Astronomy, University of Michigan, Ann Arbor, MI 48109; 
michevan@umich.edu, jbregman@umich.edu, suzacoll@umich.edu, cmullis@umich.edu}
\altaffiltext{2}{Department of Astronomy, University of California, Berkeley, CA 94720}

\begin{abstract}
Clusters of galaxies provide a closed box within which one can determine the chemical evolution of the gaseous baryons with cosmic time.  We studied this metallicity evolution in the hot X-ray emitting baryons through an analysis of XMM-Newton observations of 29 galaxy clusters in the redshift range 0.3 $<$ z $<$ 1.3.  Taken alone, this data set does not show evidence for significant evolution.  However, when we also include a comparable sample of 115 clusters observed with Chandra (Maughan et al. 2008) and a lower redshift sample of 70 clusters observed with XMM at z $<$ 0.3 (Snowden et al. 2008), there is definitive evidence for a decrease in the metallicity. This decrease is approximately a factor of two from z = 0 to z $\approx$ 1, over which we find a least-squares best-fit line $Z(z) / Z_{\odot} = (0.46 \pm 0.05) - (0.38 \pm 0.03)z$. The greatest uncertainty in the evolution comes from poorly constrained metallicities in the highest redshift bin.

\end{abstract}


\section{Introduction}

	As the largest gravitationally bound structures in the universe, galaxy clusters are thought to arise from the very first long-wavelength density perturbations in the aftermath of the big bang. Thus, measurements of observable parameters such as the cluster mass function are valuable for cosmology. On smaller scales, scaling relations between observable cluster properties such as luminosity, temperature, mass, and metal abundance are generally understood in terms of hierarchical formation models (e.g., De Lucia et al. 2005, Croton et al. 2006), so it is important to understand evolution and structure in these relations.
	
	The iron abundance of the intracluster medium (ICM) is one measurement with a number of implications. This parameter reflects the star-formation history of the galaxy cluster, since the ICM is believed to be formed from primordial gas and enriched by metals ejected by supernovae from galaxies embedded in the cluster (Matteucci \& Vettolani 1988). In nearby galaxy clusters (z $\lesssim$ 0.4), the metallicity remains close to the current canonical value of 0.3 $Z_{\odot}$ (Mushotzky \& Lowenstein 1997; Ettori, Allen, \& Fabian 2001), and the metallicity should be close to zero at very high redshifts when clusters first form, but little is known about the metallicity in the intervening times. Understanding the evolution of this parameter enables us to directly measure the evolution of star formation in cluster galaxies, which provides useful constraints for models of galaxy formation and supernova feedback (e.g., Saro et al. 2006, Romeo et al. 2006). It would be particularly useful to observe a turnover in the ICM metallicity which might represent a transition to the heavy star-formation epoch of galaxy clusters (e.g., Nagashima et al. 2005). 
	
	In addition to enriching the ICM with iron, supernovae are thought to introduce nongravitational heat to the ICM (Pipino et al. 2002) and therefore to produce deviations from self-similar scaling relations, although their relative importance compared to AGN heating is still unknown (e.g., Cavagnolo et al. 2005). This nongravitational heating appears to play a very important role in galaxy cluster dynamics and in the evolution of scaling relations (Vikhlinin et al. 2002; Ettori et al. 2004; Pratt et al. 2006; Maughan et al. 2006). 
	
	Measuring iron abundance evolution has historically posed a challenge to observers because of the difficulty of obtaining good spectra for distant clusters and the sizable errors on each measurement of abundance due to insufficient photons. Previous studies have attempted to make this measurement and have detected no significant evolution out to z $\sim$ 0.4 (Mushotzky \& Lowenstein 1997; Matsumoto et al. 2000, Snowden et al. 2008).
				
	A few studies have explored the evolution of the ICM iron abundance in clusters at z $>$ 0.4. The most comprehensive of them is a recent measurement (Maughan et al. 2008) of this evolution using the Chandra X-ray Observatory for 115 clusters from 0.1 $\leq$ z $\leq$1.3 (building on an earlier smaller study by Balestra et al. 2007).  They find evidence for evolution of the iron content comprising a decrease of about a factor of two over the redshift range. Another older measurement using 56 of these Chandra clusters (four of which also include observations with XMM-Newton) uses a slightly different spectral extraction technique and finds the same evolution, in contrast to an earlier, less robust result (Tozzi et al. 2003) which found no such evolution. These two papers represent the most direct observation of the predicted variation in iron abundance over time. The observations also match reasonably well with theoretical predictions of iron abundance evolution (Ettori 2005, Borgani et al. 2008, Fabian et al. 2008)
	
	We attempt to examine the global picture of iron abundance in galaxy clusters, using the XMM-Newton results in Snowden et al. (2008) for galaxy clusters with $z < 0.3$ and using the Chandra results in Maughan et al. and Balestra et al. for clusters at higher redshift. To supplement the high-redshift data, we also include our analysis of targeted XMM-Newton  observations of 29 galaxy clusters at $z > 0.3$. In $\S$ 2, we will discuss the sample selection and the quality of this latter data set. We will present the background analysis in $\S$ 3 and the spectral extraction in $\S$ 4. In $\S$ 5, we present the results of our metallicity analysis, and we discuss these results and their implications in $\S$ 6. Unless otherwise stated, our results are derived from the 0.3-10 keV energy range, and errors are quoted at the 1$\sigma$ confidence level. 

\section{Sample}

Our sample consists of 29 galaxy clusters at z $>$ 0.3 observed by the XMM-Newton satellite. The XMM-Newton satellite (Jansen et al. 2001) is well suited for this sort of spectral analysis. The satellite features the largest collecting area of any imaging X-ray detector, allowing us to collect as many X-ray photons as possible. For each target, we sought data from the EPIC (European Photon Imaging Camera) MOS1 and MOS2 (Multi-Object Spectrometers) CCDs and the EPIC PN detector. 

Our sample comprises all the XMM cluster observations with data available in the public archive\footnote{http://xmm.esac.esa.int/external/xmm\_data\_acc/} as of January 30, 2006 which were not critically impaired by high X-ray background. This yields a distribution of clusters from z=0.3 to z=1.3. In Fig. 1, we present the redshift distribution of our sample of 29 objects; the clusters are fairly uniformly distributed across redshift space until z $\sim$ 0.8. The cluster redshifts are provided by the NASA Extragalactic Database\footnote{http://nedwww.ipac.caltech.edu/} (NED). 

The data were first processed using the XMM-SAS (Science Analysis Sub-system; Watson et al. 2001) software, version 7.0. This standard reduction procedure corrects for instrumental artifacts and high-energy flaring effects, and produces a calibrated event file for further reduction and analysis.

In Table 1 we present the list of XMM-Newton observations. In this table, source counts are defined as the number of photons detected within the extraction radius (defined in section 3) after subtracting the background emission from the region. We note that the source counts do not comprise all of the photons expected from the cluster; to account for photons that fall outside of the source aperture, we fit a $\beta$-model (King 1962) to the emission (assuming a core radius of 250 kpc and a slope $\beta$ = 2/3) and integrate to infinite radius over the model to estimate the total emission.

\begin{deluxetable}{ccccccccc}
\tabletypesize{\scriptsize}
\tablewidth{0pt.}
\tablehead{\colhead{Cluster Name} & \colhead{R.A. (J2000)} & \colhead{Dec (J2000)}& \colhead{z}& \colhead{Obs. ID}& \colhead{Exp. Time (ks)}& \colhead{N$_d$}& \colhead{Extract. Radius (")}& \colhead{Source Counts}}
\tablecaption{\small XMM Cluster Sample}
\startdata
RXCJ1132-1955&11:31:52.0&-19:55:28&0.307&0042341001& 31.6&3& 259&  33143\\
CL0500-24&05:01:01.1&-24:25:23&0.320&0110870101& 67.6&3& 173&   5941\\
A1722&13:20:16.4&+69:59:43&0.328&0150900101& 20.4&3& 147&   6750\\
MS1208.7+3928&12:11:09.3&+39:11:34&0.340&0112190201& 36.3&3& 115&   3708\\
RXJ1532.9+3021&15:32:49.0&+30:21:15&0.345&0039340101& 33.1&3& 125&  41256\\
CL0024+17&00:26:33.1&+17:08:55&0.390&0050140201&131.5&3& 115&  13204\\
A0851&09:43:06.9&+46:59:34&0.407&0106460101&124.4&3& 160&  27242\\
RXJ2228+2037&22:28:29.8&+20:36:27&0.412&0147890101& 68.7&3& 188&  39134\\
MS0302.5+1717&03:05:22.2&+17:29:12&0.425&0112190101& 33.1&3&  91&   2219\\
RXJ1347-1145&13:47:30.8&-11:45:12&0.451&0112960101& 95.4&3& 160&  81289\\
RXJ0505.3-2849&05:05:24.0&-28:48:25&0.509&0111160201& 88.7&3&  65&   1265\\
CL0016+16&00:18:30.0&+16:25:27&0.541&0111000101& 54.7&2& 147&  29464\\
RXJ1354.3-0222&13:54:12.7&-02:21:53&0.546&0112250101& 70.1&3& 115&   3459\\
MS0451.6-0305&04:54:15.0&-03:00:18&0.550&0205670101& 79.0&3& 125&  32904\\
RXJ0847.1+3449&08:47:16.5&+34:49:22&0.560&0107860501&106.9&2&  91&   1858\\
MS2053.7-0449&20:56:18.6&-04:38:34&0.583&0112190601& 42.2&3&  77&   1472\\
RXJ0337.6-2522&03:37:48.8&-25:21:49&0.585&0107860401& 59.6&3&  47&   1129\\
RXJ1205.9+4429&12:05:46.3&+44:29:08&0.592&0156360101& 66.4&3&  44&    872\\
RXJ1120.1+4318&11:20:00.9&+43:18:15&0.600&0107860201& 53.0&3&  91&   5922\\
CL1008.7+5342&10:08:40.5&+53:42:03&0.600&0070340201& 55.5&3&  71&    873\\
RXJ1334.3+5030&13:34:15.9&+50:31:19&0.620&0111160101&108.9&3&  91&   6836\\
CL1342.8+4028&13:42:43.1&+40:28:38&0.699&0070340701& 94.9&3&  71&   2150\\
WARP1517+3127&15:17:50.3&+31:27:43&0.744&0150680101& 84.0&3&  97&   1367\\
MS1137.5+6625&11:40:29.6&+66:07:49&0.782&0094800201& 52.6&3&  98&   3437\\
MS1054.4-0321&10:56:55.5&-03:37:26&0.834&0094800101& 71.9&3&  83&   7989\\
CL1324+3011&13:24:55.4&+30:11:10&0.859&0025740201&104.0&3&  56&    803\\
CL1604+4304&16:04:25.7&+43:04:46&0.895&0025740401& 41.6&3&  48&    218\\
CL1415.1+3612&14:15:05.9&+36:11:42&1.030&0148620101& 42.3&3&  61&   1229\\
CL1252.9-2927&12:52:59.4&-29:27:13&1.237&0057740301&186.6&3&  65&   2188\\
\enddata
\label{table:sample}
\tablecomments{The Obs. ID is a unique identifier given to every observation taken with XMM-Newton. The exposure time represents the combined amount of ``clean" integration time for the three telescope cameras, after discarding photons obtained during flares or other periods of unusually high X-ray background. N$_d$ indicates the number of detectors which observed this cluster; in most cases, all three detectors were active, but CL0016+16 was observed without the M2 camera and RXJ0847.1+3449 was observed without the PN camera. Extract. Rad. represents the radius of the photometric aperture we centered around the centroid of the cluster emisssion. This radius was determined by optimizing the signal-to-noise ratio, as described in $\S$ 3. Finally, the last column indicates the number of source counts detected for each cluster (summed over all detectors).}
\end{deluxetable}

\section{Background Subtraction and Extraction Radius}

Before we derive the galaxy cluster spectra from the energy distribution of the incident X-ray photons on the three detectors, we must first consider a number of sources of photon contamination. The most prominent of these are the X-ray background (XRB) and resolved point sources, and the detector background. To mitigate these effects, we attempted to categorize their contributions and remove them from our images, by removing point sources and then using a standard in-field background subtraction technique. For point sources, we identified and removed them automatically using the XMM-SAS task EBOXDETECT, and verified the results visually. 

We selected the in-field subtraction technique for our background analysis. The in-field background subtraction technique involves selecting two regions in an image - a circular ``source" region including most of the galaxy cluster photons, and an annular ``background" region outside of the source which includes a representative sample of the XRB and particle background in the image. The spectrum of the background region is then computed, and subtracted from the source region to produce a cleaned spectrum that represents our best estimate of the ``true" galaxy cluster spectrum. We are able to use this method because galaxy clusters at these distances generally only comprise an arcminute or two in diameter, whereas typical variations in the unresolved XRB and in the detector response occur over larger angular scales. Thus, we do not correct for detector response effects until after selecting our source and background regions.

The source region should be defined so as to include as many photons as possible from the galaxy cluster while minimizing contamination from background photons. We account for photons outside the source aperture later in our analysis. The region selection methodology must also be uniform across all our observations, and it must be reproducible. To accomplish all of these goals, we used a signal-to-noise optimization procedure based on the X-ray surface brightness profile of the galaxy cluster. By computing the signal-to-noise ratio for concentric circular source apertures comprising eighty logarithmically spaced radii, we were able to select the radius that maximized this ratio for each individual cluster.

\section{Spectral Analysis}

After selecting source and background regions, we extracted the spectra from these regions. We analyzed the spectra with XSPEC v11.3.1 (Arnaud 1996), and fitted parameters to data from the three detectors jointly using a single-temperature MEKAL model (Kaastra 1992; Liedahl et al. 1995) modified to account for interstellar absorption (McCammon \& Sanders 1990) using known Galactic neutral hydrogen column densities along the line of sight (Dickey \& Lockman 1990). We re-binned our spectral files to include at least 20 counts per bin. The energy range was 0.3-10 keV. 

In the spectral analysis, we fixed the redshift and N$_H$, and left temperature, metallicity, and normalization as free parameters for a Chi-squared minimization fit across all three detectors simultaneously. After spectral fitting, we verified that the reduced $\chi^2$ was on the order of unity for all of our fits. 
We used a standard detector model to weight received energy by the response across the detector. In one case (A1722), however, the S/N was too poor to produce a good fit after this vignetting correction, so we used a flat detector model for this cluster in our spectral fitting.

Finally, we normalized our iron abundances to the solar abundances in Anders and Grevesse (1989), to facilitate comparison with results of other iron abundance papers. The results of the spectral fits are presented in Table 2, and the iron abundances of our 29 clusters are displayed against redshift in Fig. 2. We also compared the uncertainty in iron abundance in our analysis of each cluster with the number of source photons (Fig. 3). If the uncertainty were dominated by photon statistics, it should scale linearly with the square root of the number of source photons. We find such a scaling for clusters with at least $\sim$ 2000 photons; below that, background effects and other sources of error seem to dominate.

\begin{deluxetable}{ccccc}
\tabletypesize{\scriptsize}
\tablewidth{0pt.}
\tablehead{\colhead{Name} & \colhead{z} & \colhead{T$_X$}& \colhead{Z}& \colhead{N$_H$ (10$^{20}$ cm$^{-2}$)}}
\tablecaption{\small Spectral Fits}
\startdata
RXCJ1132-1955&0.307&$  7.58^{+   0.21}_{   -0.21}$&$  0.23^{+   0.04}_{   -0.04}$&       4.23\\
CL0500-24&0.320&$  4.07^{+   0.28}_{   -0.25}$&$  0.32^{+   0.10}_{   -0.10}$&       2.40\\
A1722&0.328&$  6.20^{+   0.36}_{   -0.33}$&$  0.24^{+   0.09}_{   -0.08}$&       1.61\\
MS1208.7+3928&0.340&$  5.31^{+   0.43}_{   -0.40}$&$  0.33^{+   0.14}_{   -0.15}$&       1.89\\
RXJ1532.9+3021&0.345&$  5.15^{+   0.09}_{   -0.08}$&$  0.31^{+   0.02}_{   -0.03}$&       2.11\\
Cl0024+17&0.390&$  3.63^{+   0.16}_{   -0.10}$&$  0.28^{+   0.06}_{   -0.05}$&       4.09\\
A0851&0.407&$  5.32^{+   0.15}_{   -0.14}$&$  0.20^{+   0.04}_{   -0.04}$&       1.19\\
RXJ2228+2037&0.412&$  7.85^{+   0.19}_{   -0.19}$&$  0.20^{+   0.03}_{   -0.04}$&       4.78\\
MS0302.5+1717&0.425&$  8.37^{+   1.23}_{   -0.96}$&$  0.69^{+   0.33}_{   -0.30}$&       1.09\\
RXJ1347-1145&0.451&$ 10.68^{+   0.13}_{   -0.12}$&$  0.27^{+   0.02}_{   -0.02}$&       4.67\\
RXJ0505.3-2849&0.509&$  2.65^{+   0.30}_{   -0.26}$&$  0.05^{+   0.15}_{   -0.05}$&       1.50\\
CL0016+16&0.541&$  9.48^{+   0.28}_{   -0.26}$&$  0.21^{+   0.04}_{   -0.04}$&       4.09\\
RXJ1354.3-0222&0.546&$  4.79^{+   0.71}_{   -0.57}$&$  0.20^{+   0.20}_{   -0.19}$&       3.40\\
MS0451.6-0305&0.550&$  7.97^{+   0.22}_{   -0.19}$&$  0.21^{+   0.03}_{   -0.03}$&       5.65\\
RXJ0847.1+3449&0.560&$  4.30^{+   0.59}_{   -0.48}$&$  0.24^{+   0.19}_{   -0.18}$&       3.20\\
MS2053.7-0449&0.583&$  5.10^{+   0.71}_{   -0.55}$&$  0.29^{+   0.24}_{   -0.21}$&       4.68\\
RXJ0337.6-2522&0.585&$  4.23^{+   0.53}_{   -0.46}$&$  0.38^{+   0.36}_{   -0.28}$&      0.990\\
RXJ1205.9+4429&0.592&$  2.83^{+   0.34}_{   -0.26}$&$  0.37^{+   0.28}_{   -0.22}$&       1.28\\
RXJ1120.1+4318&0.600&$  5.23^{+   0.25}_{   -0.23}$&$  0.47^{+   0.09}_{   -0.09}$&       2.10\\
CL1008.7+5342&0.600&$  3.13^{+   0.44}_{   -0.20}$&$  0.32^{+   0.36}_{   -0.27}$&      0.766\\
RXJ1334.3+5030&0.620&$  5.83^{+   0.42}_{   -0.37}$&$  0.09^{+   0.12}_{   -0.04}$&       1.05\\
CL1342.8+4028&0.699&$  3.83^{+   0.34}_{   -0.30}$&$  0.30^{+   0.19}_{   -0.17}$&      0.792\\
WARP1517+3127&0.744&$  2.99^{+   0.53}_{   -0.37}$&$  0.40^{+   0.43}_{   -0.28}$&       1.87\\
MS1137.5+6625&0.782&$  6.56^{+   0.76}_{   -0.64}$&$  0.22^{+   0.14}_{   -0.15}$&       1.08\\
MS1054.4-0321&0.834&$  8.06^{+   0.42}_{   -0.39}$&$  0.22^{+   0.07}_{   -0.08}$&       3.87\\
CL1324+3011&0.859&$  4.24^{+   1.05}_{   -0.50}$&$  0.35^{+   0.56}_{   -0.33}$&       1.14\\
CL1604+4304&0.895&$  5.04^{+   2.51}_{   -1.58}$&$  0.00^{+   1.56}_{   -0.00}$&       1.31\\
CL1415.1+3612&1.030&$  5.47^{+   0.75}_{   -0.62}$&$  0.53^{+   0.32}_{   -0.29}$&       1.06\\
CL1252.9-2927&1.237&$  5.00^{+   0.85}_{   -0.51}$&$  0.16^{+   0.15}_{   -0.16}$&       5.90\\
\enddata
\tablecomments{Each cluster is identified by its name (see Table 1) and presented in ascending redshift order. T$_X$ is the average temperature of the cluster ICM in keV with 1$\sigma$ errors. The abundance is given in solar abundances with 1$\sigma$ errors. N$_H$ is the column density. }\label{table:summary}
\end{deluxetable}

\section{Comparison of High-Redshift XMM and Chandra Results}

We checked the results of our spectral analysis against those in the literature (e.g., Lumb et al. 2004, Bartlett et al. 2001, Novicki et al. 2002, Balestra et al. 2007, Snowden et al. 2008, Maughan et al. 2008). In all of these cross-checks, our temperature and iron abundance values generally fall within 1 $\sigma$ of the published values, except for a discrepancy with the iron abundance as measured by Chandra. This discrepancy has been noted by others (e.g. Snowden et al. 2008), and a new calibration update (CIAO CALDB 4.4.1) has been released recently that may correct the discrepancy if applied to the Chandra data. We also see a slight trend towards measuring a lower temperature and abundance in our XMM sample compared to Chandra as redshift decreases (Fig. 4), but the trend is only marginally significant and possibly related to the calibration issue discussed above: the temperature ratios are inconsistent with unity with a $\chi^2$ of 17.4 (not significant; P $<$ 0.10, 10 d.o.f.), while the iron abundance ratios are inconsistent with unity with a $\chi^2$ of 35.1 (significant; P $<$ 0.001, 10 d.o.f.).

To verify that our spectral analysis procedure has no significant systematic biases, we compared the eleven $z > 0.3$ clusters that were observed by both XMM (this paper) and Chandra (Maughan et al.) to see if the uncertainties were behaving as expected. In Fig. 5, we compare the ratio of exposure times to the ratio of iron abundance uncertainties for these eleven clusters. For photon-dominated noise, there should be a simple $t^{-1/2}$ relation between exposure time and uncertainty for each instrument, but the two telescopes have different sensitivities and collecting area, so the exposure times are not exactly equivalent. We used the HEASARC WebSpec utility\footnote{http://heasarc.gsfc.nasa.gov/webspec/webspec.html} to compute the expected photon count for each instrument given an observation of equal time for the same galaxy cluster, and we estimate that XMM-Newton should collect about three times as many photons per second as Chandra. The dashed line in Fig. 5 represents this relation. The quality of the XMM data matches the expectations well, except for the two points on in the upper left. Those represent the highest-redshift clusters: the highest point is a cluster with a problematic fit (Z = 0) to the Chandra data and therefore an artificially low uncertainty in its iron abundance.

Following Maughan et al. (2008), we also binned our sample into redshift intervals, using the same bins as in their work.  The first bin comprised the five XMM clusters with $0.3 < z < 0.37$. The second bin comprised five clusters from $0.37 \le z < 0.46$. The third bin contained four clusters from $46 \le z < 0.55$, the fifth nine clusters from $0.55 \le z < 0.75$, and the last six clusters were placed into the $z > 0.75$ bin. We used a different method for averaging over the clusters in each bin, however. Maughan et al. computed their iron abundance using a joint spectral fit to all of the clusters in the bin with the iron abundance fixed between all the clusters. This technique has the advantage of automatically weighing the contribution of each cluster by the number of photons received from that cluster. But it also underestimates the uncertainty on the derived iron abundance: poor fits with unacceptably high $\chi^2$ are generally more sensitive to small deviations in iron abundance than better fits with a lower $\chi^2$, so these poor fits often significantly underestimate the 1$\sigma$ uncertainty, and this underestimation dominates in a joint fit and leads to an underestimation of the total uncertainty in such a fit.  For example, there are three clusters out of fifteen in the lowest-redshift bin ($0.3 < z < 0.37$) that are more than two standard deviations (including uncertainty on both the individual cluster iron abundance and the joint fit abundance) from the joint fit value; we would expect only one such cluster if the uncertainties were Gaussian. 

Therefore we computed the iron abundances in each bin by averaging the iron abundance of the clusters in the bin, weighing each value by the inverse of the variance on the iron abundance. We also assigned larger uncertainties to clusters with failed spectral fits based on detailed explorations of the parameter space instead of using the default XSPEC value. If the uncertainty is dominated by insufficiency of photons, our weighing scheme should have the same effect as weighing by the number of photons, but without the problem of poor fits artificially decreasing the uncertainty. We also recalculated the iron abundance and uncertainties for the Maughan et al. data, using their data for each cluster in their Table 3. This did not significantly change the average iron abundances, but it did increase the uncertainties by about 50\%. As a final check, we also independently computed average iron abundances and uncertainties using the median value as the center and the abundances containing 50\% of the clusters as the uncertainties; if the errors are randomly distributed, this method should be unaffected by the magnitude of the uncertainties, and we found essentially the same behavior with these bins as with our weighted bins.

We also include in our analysis XMM-Newton measurements of iron abundance for a sample of 70 clusters (Snowden et al. 2008), of which 68 have $z < 0.3$ (one of the other two is also included in our sample). For these clusters, instead of measuring a single emission-weighted iron abundance, they measured the iron abundance in successive annuli around the core. We converted these measurements into an emission-weighted average by weighing the iron abundance of each annulus by the region's flux and angular size. For the one cluster in both our sample and their paper (RXJ1347-1145), we measured $Z = 0.27 \pm 0.02 Z_{\odot}$, and using this weighting scheme we derive a measurement for Snowden et al. of $Z = 0.35 \pm 0.04 Z_{\odot}$. We binned these measurements using the same bins as Maughan et al. for $0.1 < z < 0.3$ and added a few additional bins at low redshift. 

\section{Results}

We examine iron abundance as a function of redshift (Fig. 2) for the clusters in our sample. In the initial plot, no significant trend is visible, mainly due to the large uncertainty associated with each individual cluster. We attempted to characterize the correlation between iron abundance and redshift using the Spearman's Rank Coefficient, which provides a statistical measurement of the likelihood of a correlation between two sets of variables (in our case, redshift and iron abundance). For these two parameters we find a coefficient of $r_S$ = 0.007, corresponding to a detection of evolution with a p-value of 0.97. Usually, a detection is only considered statistically significant if the p-value is very low (less than 0.05 or 0.01), so our result is consistent with no evolution.

To characterize the evolution more precisely, we binned our sample into redshift intervals as described above and compared to data from the Chandra X-ray Observatory (Maughan et al.) and to lower-redshift data from XMM-Newton (Snowden et al.). We present the resulting plot in Fig. 6. For our binned $z > 0.3$ XMM data, we find a Spearman's rank coefficient of $r_S$ = -0.50, corresponding to a detection of evolution (decreasing abundance) with a p-value of 0.39. This is not a significant detection, and furthermore the rank coefficient does not consider the uncertainty on the iron abundance measurement. We account for these errors by fitting a line using Chi-squared minimization. The best-fit is: $Z / Z_{\odot}$ = (0.34 $\pm$ 0.10) - \linebreak (0.18 $\pm$ 0.22)z. Note that the 1$\sigma$ error on the slope of the line is greater than the actual slope. We plot this line and the 1$\sigma$ deviations from it with the data. 

Before considering all three datasets together, we note that the discrepancy mentioned above between the Chandra and XMM iron abundances could also obviously affect a measurement of abundance using both datasets. The discrepancy appears to affect primarily the low-redshift iron abundances, so its effect will presumably be smaller at high redshift. We consider two limiting cases: the case where we ignore the discrepancy, and the case where we assume a constant discrepancy of a factor of 0.8 across the redshift space (scaling by this factor brings the Chandra and our XMM data into agreement at the $z = 0.3$ bin).

Considering all three datasets simultaneously, in the former case even with our expanded uncertainties we can strongly rule out a zero evolution model: for a constant iron abundance of 0.4 $Z_{\odot}$ and eighteen degrees of freedom, we find a $\chi^2$ of 173.6, which corresponds to much better than a 99.9\% confidence level. This is consistent with the evolutionary trend described by Maughan et al.: computing the least-squares best-fit line (Z(z)) and 1$\sigma$ deviations to the data in Fig. 6, we find $Z / Z_{\odot}$ = (0.46 $\pm$ 0.05) - (0.33 $\pm$ 0.03)z. The Spearman's rank coefficient is $r_S$ = -0.92, with a p-value of $6.6\times10^{-5}$, which is indeed statistically significant (at the 4$\sigma$ level). 

We note, however, that the nature of the evolution appears to be driven significantly by the highest-redshift data points in Fig. 6. With Chandra, this point contains weighted abundances from eleven clusters, but four of them have problematic spectral fits, which are noted as zero metallicity in Maughan et al. These clusters bring down the average iron abundance in the bin, and could introduce a bias into searches for evolution since more distant galaxy clusters typically have fewer photons and thus more of these problematic spectral fits. Indeed, if we display the median iron abundance in each bin and use the range containing 50\% of the iron abundances (Fig. 7), the significance of the evolution at z $>$ 0.3 diminishes.

Scaling the Chandra data by a factor of 0.8 and considering a joint fit to these data and all the XMM data  yields a very similar best-fit line: $Z / Z_{\odot}$ = (0.46 $\pm$ 0.05) - (0.38 $\pm$ 0.03)z. The Spearman's rank coefficient is also essentially unchanged: $r_S$ = -0.93, with a p-value of $8.2\times10^{-5}$. Additionally, for a constant iron abundance of 0.40 $Z_{\odot}$ the $\chi^2$ is increased to 258 for eighteen degrees of freedom, ruling out zero evolution even more strongly.

\section{Discussion}

We have measured the iron abundances of the ICM in 29 galaxy clusters between 0.3 $<$ z $<$ 1.3 using archival XMM-Newton observations. These observations, while fewer in number than comparable Chandra observations, are competitive with existing Chandra data due to the greater collecting area of XMM-Newton. Using just these XMM data, we do not find significant evidence for evolution in the ICM iron abundance. However, combining these data with existing Chandra data and low-redshift XMM data, the suggestion of evolution is much stronger - a 4$\sigma$ signal, and higher than 99.9\% statistical confidence. We do note that this evolution is somewhat less impressive when defined by the median values instead of by weighted means, and could be driven in part by cooling cores in the low-redshift (z $<$ 0.3) data.

There is also a systematic difference between measurements of iron abundance in clusters common to our sample and the Chandra sample, and this difference appears to depend on redshift. This may be a calibration issue (see Snowden et al. 2008), and furthermore it could also be responsible for some of the observed metallicity gradient. Reprocessing of the Chandra data with newer calibration files may mitigate this effect.

Additionally, the observed evolution could theoretically be driven by the growth of cool cores in galaxy clusters, since these increase the flux in the center of the cluster where the iron abundance is greatest, therefore increasing the overall emission-weighted iron abundance. However, Maughan et al. looked into this issue by explicitly excluding the central region ($r < 0.15 r_{\text{vir}}$) from their analysis and they find the same trend and the same magnitude of decline

As Figs. 6 and 7 suggest, the paucity of high-redshift clusters drives the uncertainty in the iron abundance evolution; we simply lack sufficient accurate measurements of iron abundance in clusters beyond $z \sim 0.7$. Unfortunately, in enrichment models such as Ettori (2005), the metallicity evolution follows closely to the star formation history, and so this higher-redshift region contains most of the evolution. Thus, while we can say confidently that iron abundance increases from $z \sim 1$ to the present, we cannot constrain the iron abundance at $z \sim 1$ very well. It may become very low beyond these redshifts, or it may flatten out at about $0.1 Z_{\odot}$ or $0.2 Z_{\odot}$ as the result of a previous pre-enrichment phase. More high-quality observations of high-redshift clusters should enable significant reduction of the uncertainties on iron abundance at high redshift and distinguish between these models. This will allow for much more sensitive constraints on the evolution of iron abundance and the associated formation history of galaxy clusters.

\acknowledgements
This work is based on observations obtained with XMM-Newton, an ESA science mission with instruments and contributions directly funded by ESA Member States and NASA. This research has made use of SAOImage DS9, developed by Smithsonian Astrophysical Observatory. This research has made use of NASA's Astrophysics Data System. 

{\it Facilities:} \facility{XMM}, \facility{Chandra}

\begin{figure}
\epsscale{1.0}
\plotone{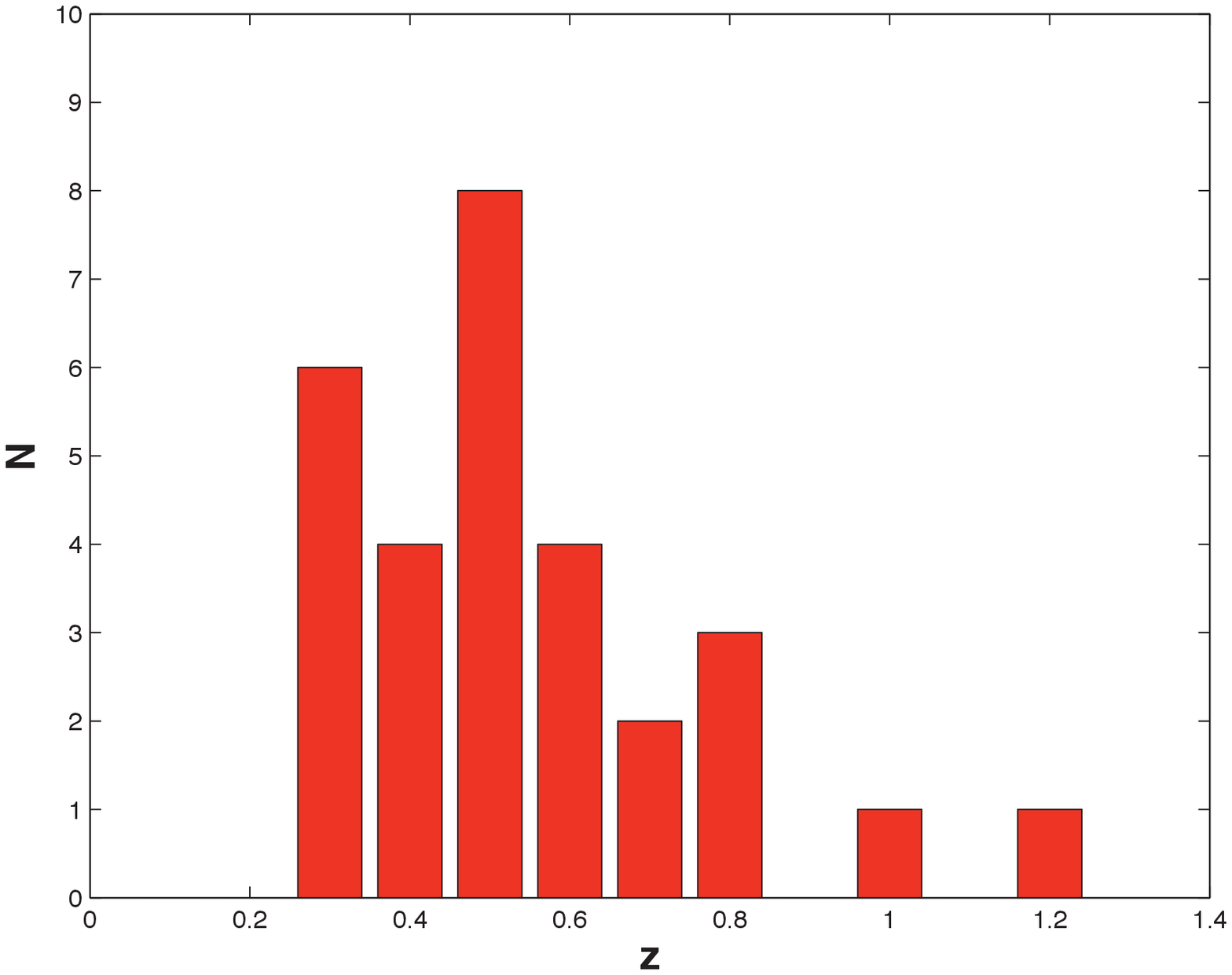}
\caption{\small Redshift histogram of the 29 galaxy clusters in our sample.}
\end{figure}

\begin{figure}
\epsscale{1.0}
\plotone{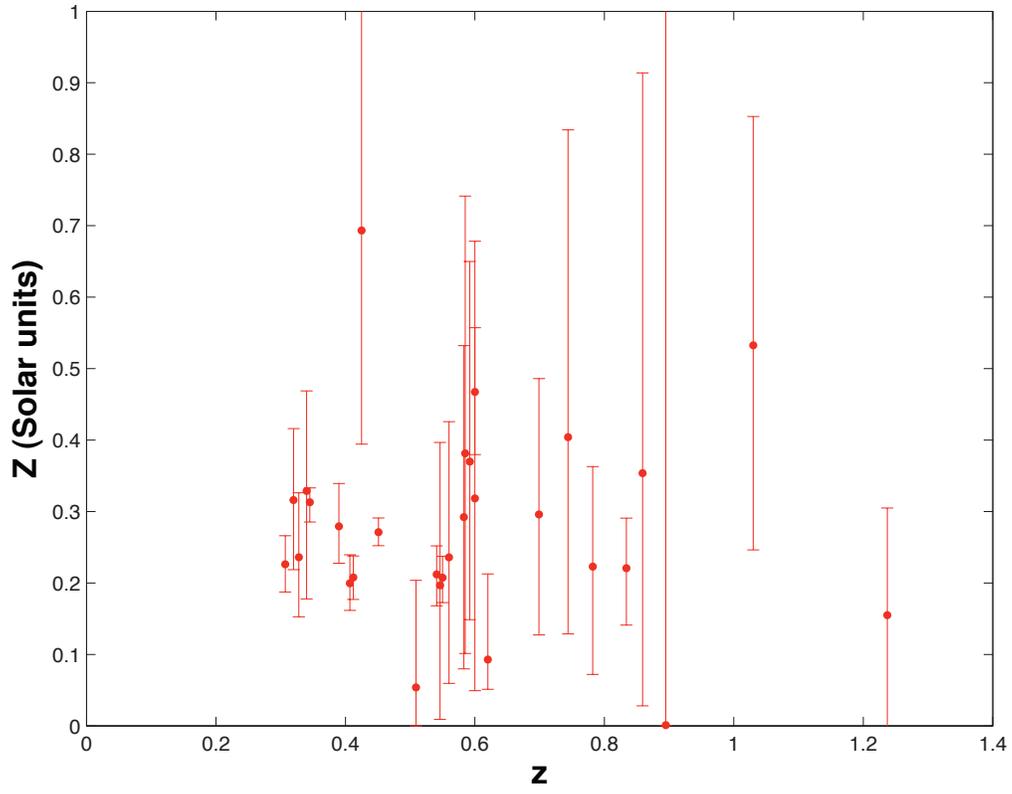}
\caption{\small  Iron abundance and 1$\sigma$ error bars as a function of redshift for our XMM galaxy cluster sample.}
\end{figure}

\begin{figure}
\epsscale{1.0}
\plotone{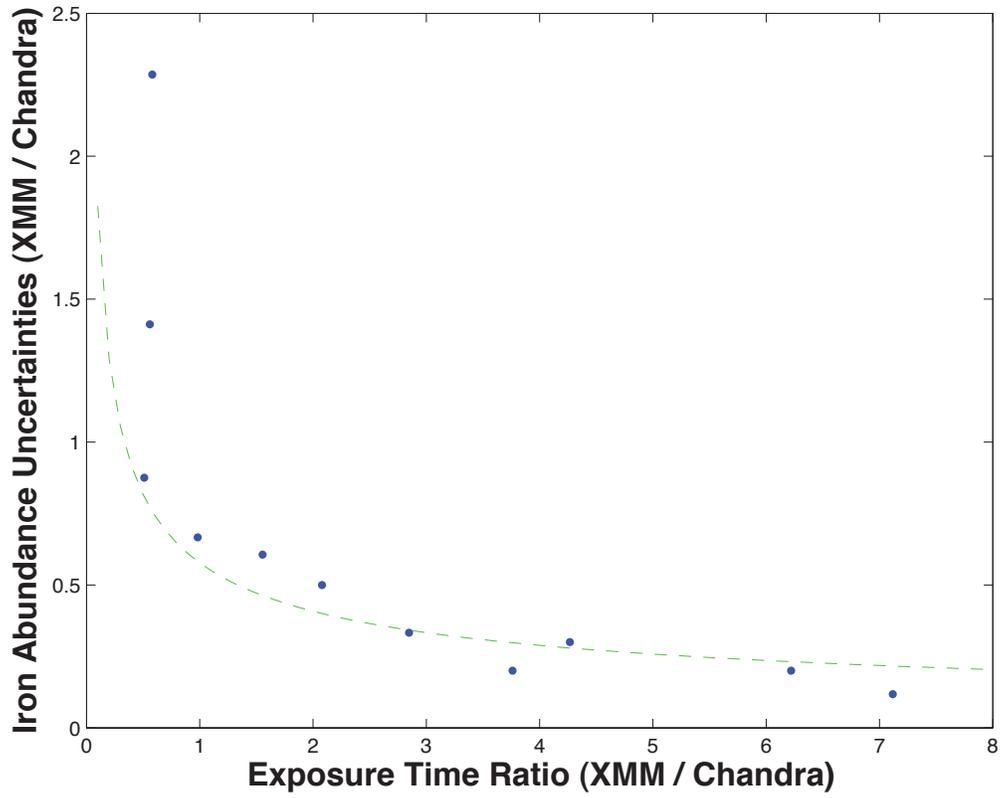}
\caption{\small Errors in iron abundance as a function of the square root of the source photon count. For N $>$ 2000, the error decreases roughly linearly, suggesting photon statistics dominate the uncertainty. Other effects seem to dominate for smaller N, such as errors from the in-field background subtraction}
\end{figure}

\begin{figure}
\epsscale{1.0}
\plotone{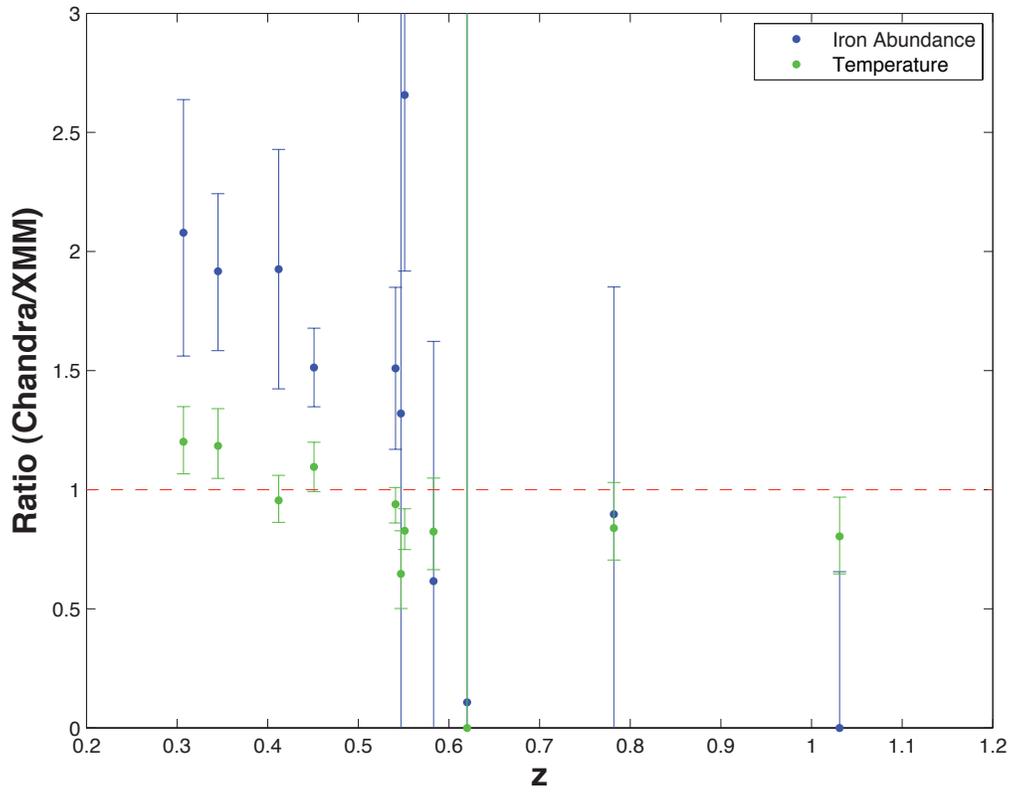}
\caption{\small Comparison of measured $T_X$ and iron abundance for eleven clusters observed by both Chandra (Maughan et al. 2008) and XMM (this paper). While the temperatures agree well, we note a discrepancy in the iron abundances at lower redshift, which may have been corrected in a newer version of the Chandra calibration code.}
\end{figure}

\begin{figure}
\epsscale{1.0}
\plotone{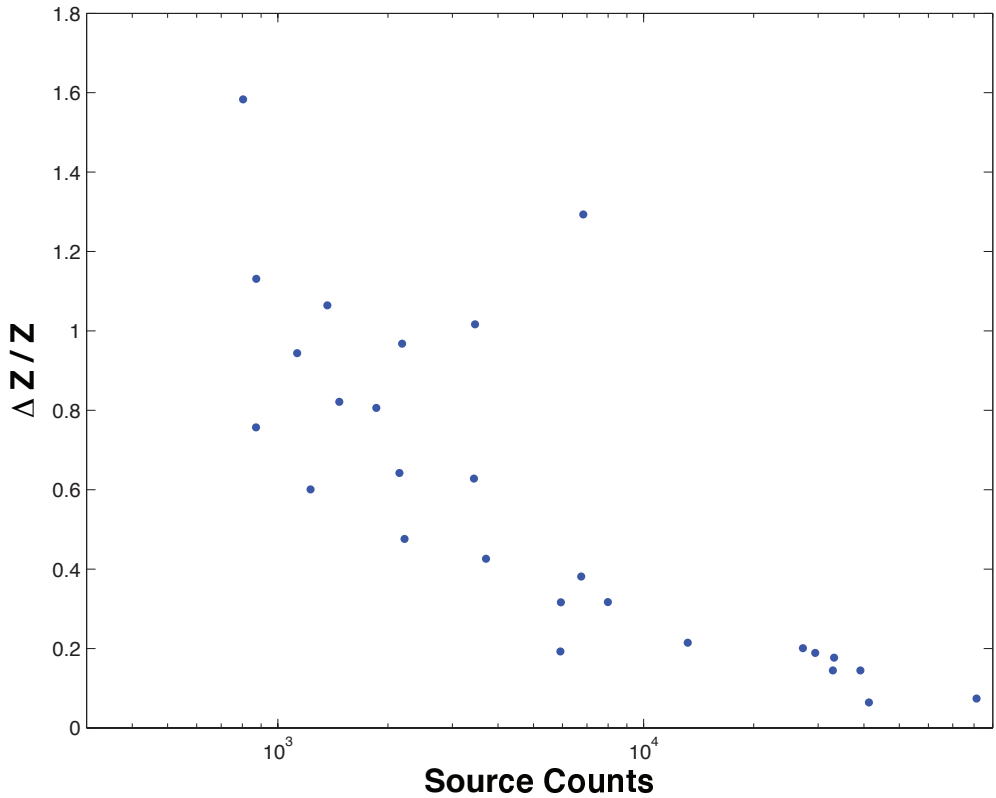}
\vspace{-0.5 in}
\caption{\small Comparison of exposure time ratio and iron abundance uncertainty for clusters observed by both Chandra (Maughan et al. 2008) and XMM (this paper). The dashed line denotes the expected relation between the two telescopes assuming photon statistics dominate the noise in each. Note that XMM-Newton can achieve smaller uncertainties with less exposure time than Chandra due to its larger collecting area.}
\end{figure}

\begin{figure}
\epsscale{1.0}
\plotone{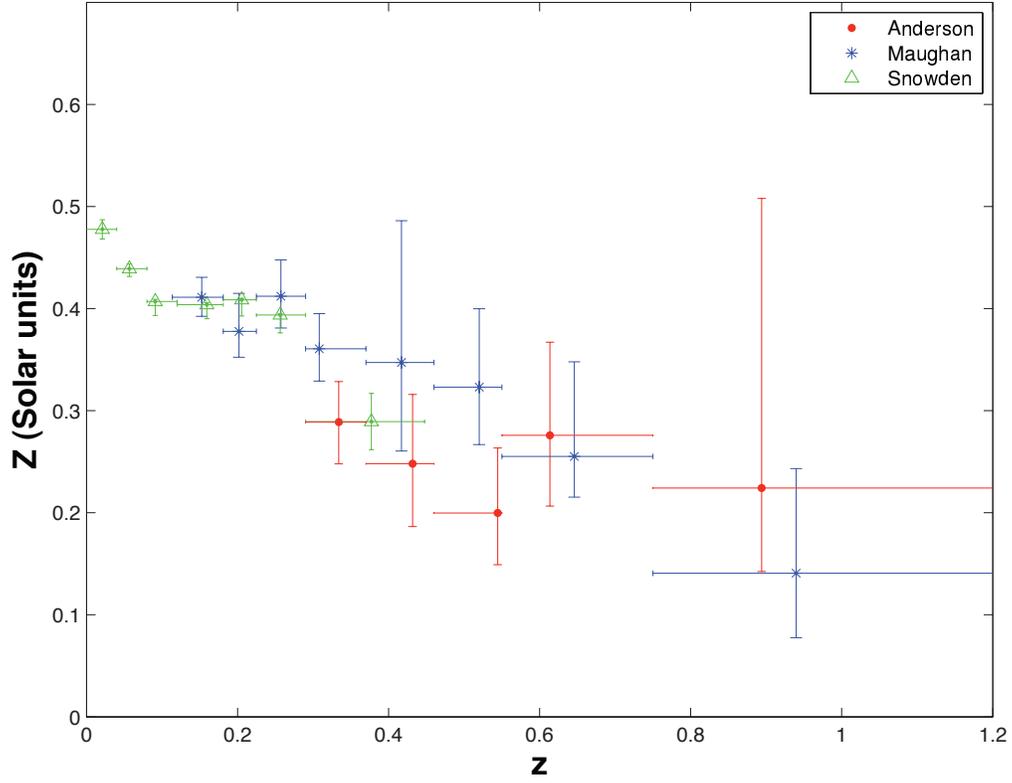}
\vspace{-0.5 in}
\caption{\small Weighted means of iron abundance as a function of redshift for our analysis (red circles), Maughan et al (blue asterisks), and Snowden et al. (green triangles). The means are obtained by averaging over the iron abundance for each individual cluster and weighing the abundance and the associated $1\sigma$ uncertainty by the uncertainty on each cluster. The least-squares best-fit to these data is $Z / Z_{\odot}$ = (0.46 $\pm$ 0.05) - (0.33 $\pm$ 0.03)z.}
\end{figure}

\begin{figure}
\epsscale{1.0}
\plotone{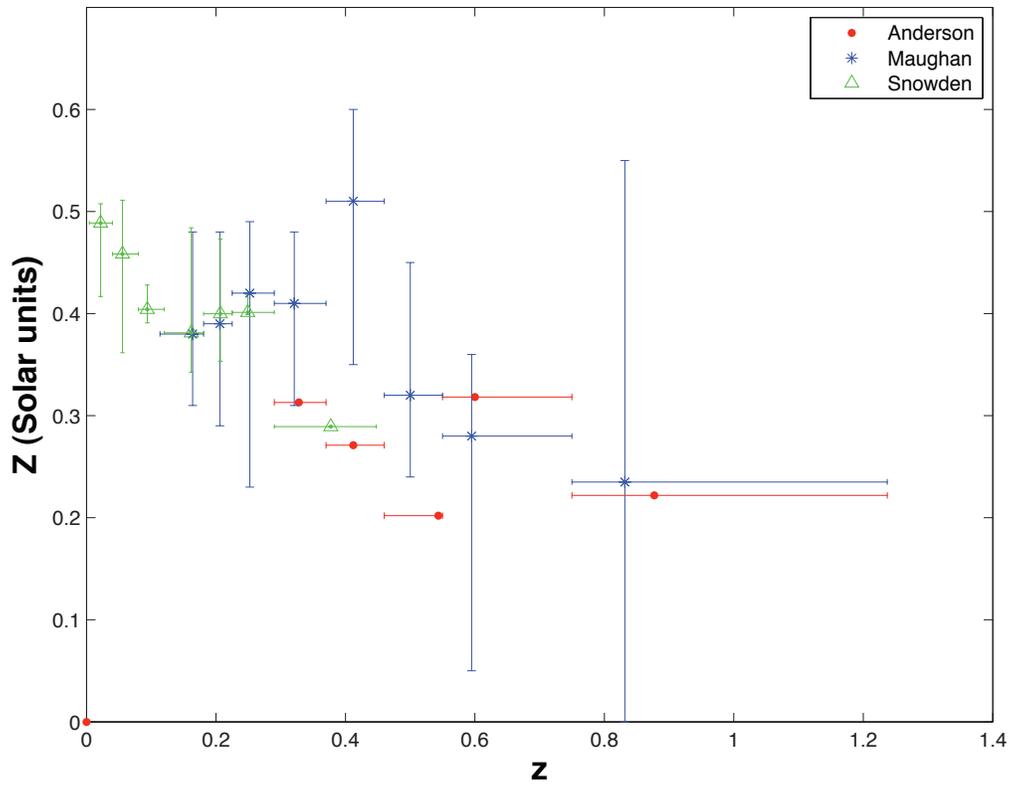}
\vspace{-0.5 in}
\caption{\small Median and quartile range in each bin for iron abundance as a function of redshift for our analysis (red circles),  Maughan et al. (blue asterisks), and Snowden et al. (green triangles). The bins in our analysis and the highest-redshift Snowden et al. bin do not contain enough clusters to make a quartile meaningful. The trend from Fig. 6 is visible here as well, but the uncertainties on each point are much larger and the constraints weaker.}
\end{figure}

\end{document}